\documentclass[twocolumn,secnumarabic,amssymb, nobibnotes, aps, prd]{revtex4}

\usepackage{graphicx}

\begin{document}

\title{Magic radioactivity of $^{252}$Cf}
%\title{Microscopic cold fission yields of $^{252}$Cf}

\author{M. Mirea$^{1}$, D.S. Delion$^{1,2}$ and A. S\u andulescu$^{2,3}$}
\affiliation{
$^{1}$National Institute of Physics and Nuclear Engineering,\\
407 Atomi\c stilor, 077125 Bucharest-M\u agurele, Romania \\
$^{2}$Academy of Romanian Scientists \\
54 Splaiul Independen\c tei, 050094 Bucharest, Romania \\
$^{3}$Institute for Advanced Studies in Physics, \\
129 Calea Victoriei, Bucharest, Romania}

\begin{abstract}
{We show that the sharp maximum corresponding to $^{107}$Mo
in the fragment distribution of the $^{252}$Cf cold fission
is actually a Sn-like radioactivity, similar to  other decay processes 
in which magic nuclei are involved, namely $\alpha$-decay and heavy 
cluster emission, also called Pb-like radioactivity. 
It turns out that the mass asymmetry degree of freedom
has a key role in connecting initial Sn with the final Mo isotopes
along the fission path. 
We suppose the cold rearrangement of nucleons within the framework
of the two center shell model, in order to compute the cold valleys
in the charge equilibrated fragmentation potential.
The fission yields are estimated by using the semiclassical
penetration approach.
We consider five degrees of freedom, namely the inter-fragment
distance, the shapes of fragments, the neck parameter and
mass asymmetry.
We found an isomeric minimum between the internal and external barriers.
It turns out that the inner cold valley of the total potential energy 
is connected to the double magic isotope $^{132}$Sn.}
\end{abstract}

\vskip1cm



\maketitle
%\date{\today}

%\newpage

\rm

\setcounter{equation}{0}
\renewcommand{\theequation}{\arabic{equation}}

%\section{Introduction}
\label{sec:intro}

The spontaneous cold rearrangement of nucleons from an initial
nucleus to two final fragments corresponds to the most favorable 
path of the cold splitting in the potential energy surface \cite{San76}.
This path, called cold valley, is related to the magicity
of one or both fission fragments. 
Thus, $\alpha$-decay is connected to the cold valley of the double 
magic nucleus $^{4}$He.
The cold valley of the double magic $^{208}$Pb is responsible for
various heavy cluster decays, in which C, O, Ne, Mg and Si are 
emitted  \cite{San80} and this is the reason why they are also called 
"magic-radioactivities".

On the other hand the production of superheavy elements is 
connected with the inverse fusion process
\cite{Oga01}, involving double magic nuclei 
$^{208}$Pb and $^{48}$Ca \cite{Gup77,Gup77a}.

Two main synthesis procedures of superheavy nuclei based on fusion 
reactions have been experimentally used:

(a) the cold-fusion at GSI Darmstadt with either $^{208}$Pb or 
$^{209}$Bi target
leading to a small excitation energy of the compound nucleus followed
by a single neutron evaporation \cite{Hof00,Hof04},
and

(b) the hot fusion with $^{48}$Ca projectile at JINR Dubna, in which the
compound nucleus is very excited hence more neutrons are evaporated
\cite{Oga01,Oga04}.

%The production of many superheavy elements with $Z\leq 118$
%(corresponding to Cf, as the last possible target) during last three 
%decades was mainly based on the existence of cold valleys
%for $^{48}$Ca in the fragmentation potential

Almost two decades ago, systematic measurements were 
performed to determine cold fission yields of $^{252}$Cf 
\cite{ham1,ham2}.
The aim of this work is to analyze these data and to show that the cold 
fission of $^{252}$Cf is strongly connected with the cold valley of the 
double magic isotope $^{132}$Sn,
although the experimental cold yields have a 
maximum corresponding to a different charge number.

In the past the cold fission of $^{252}$Cf was investigated within the 
double folding  potential method \cite{sandu96,sandu98} emphasizing the 
role of higher deformations.
In Ref \cite{gonnen91}, based on a macroscopic model by determining the tip 
distances for the exit point from the barrier for ground state deformed 
fragments, it was predicted that the major contribution in the yield 
distribution corresponds to the light fission fragment 
$A_{2}\approx$100, contradicting experimental data showing a peak at
$A_2=107$. In this study we explain the root of this discrepancy.
 
We extend the analysis performed in Ref. \cite{Del07}
to a more reliable microscopic approach to estimate the 
fission barrier, based on a new version of 
the Super Asymmetric Two Center Shell Model \cite{mirws2}.
This approach was already used to describe the fusion/fission of 
some superheavy elements \cite{superheavy} and of the dynamical effects
in fission \cite{mirws2,Mir99}. 

In this work the nuclear shape parametrization is obtained by smoothly joining
two spheroids with a third surface, given by the rotation of a circle
around the symmetry axis. This parametrization is characterized
by 5 degrees of freedom, namely the mass asymmetry, the elongation 
$R=z_2-z_1$ given by the distance between the centers of the nascent 
fragments, the  two deformations of the nascent fragments associated to 
the eccentricities $\varepsilon_{i}=\sqrt{1-b_{i}^2/a_{i}^2}$ ($i$=1,2),
and the necking characterized
by the curvature $C=S/R_{3}$ of the median surface and a asymmetry term 
given by the ratio of the semi-axis $\eta=a_{1}/a_{2}$. In this parametrization
the values $S$=1 and $S$=-1 stand for necked and swollen shapes 
in the median surface,
respectively. Within
these generalized coordinates, for necked shapes it is possible to estimate
the mass of the nascent fragment $A_2$ during the deformation.

\begin{figure}

\begin{center}
\resizebox{0.50\textwidth}{!}{
  \includegraphics{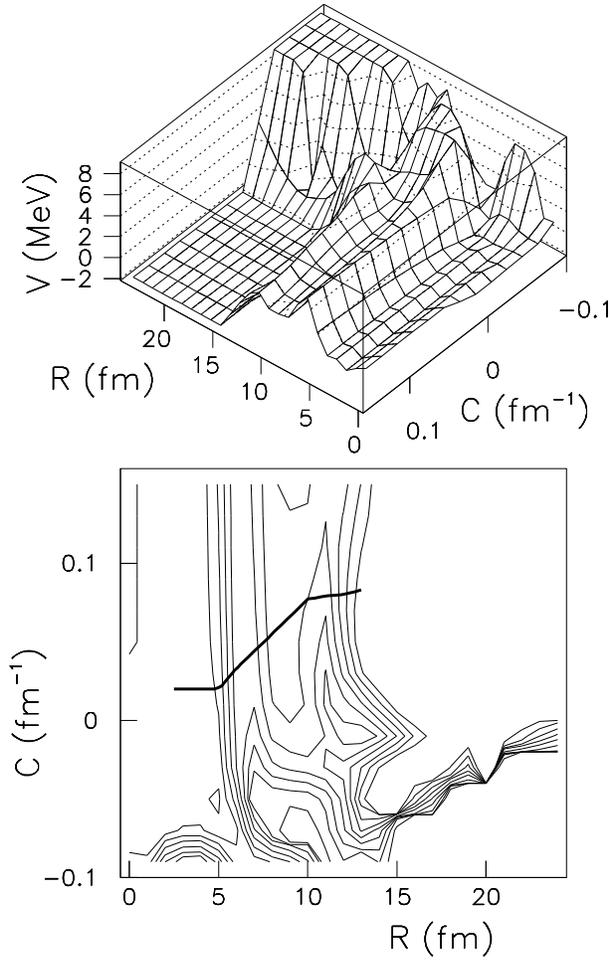}}

\caption{Deformation energy $V$ computed within the microscopic-macroscopic
method with respect $C$ (necking coordinate) and $R$ (elongation). 
In the lower plot,  the minimal action trajectory is also plotted. The step 
between two equipotential lines is 1 MeV.
 }
\label{fig1}
\end{center}
\end{figure}

\begin{figure}

\begin{center}
\resizebox{0.50\textwidth}{!}{
  \includegraphics{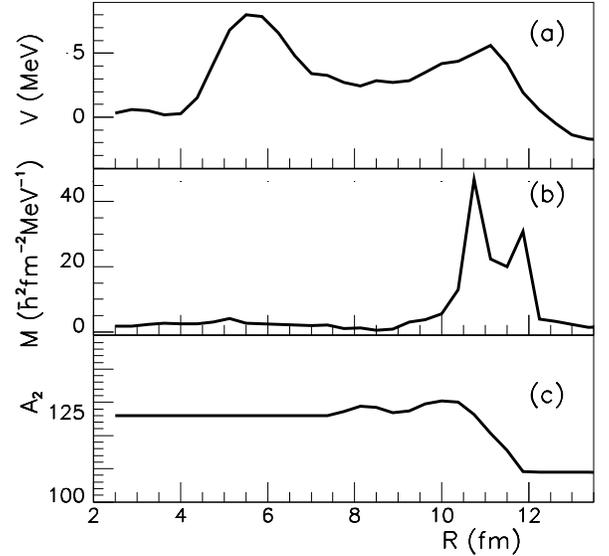}}

\caption{(a) Adiabatic fission barrier as function of the elongation $R$ along
the minimal action path. (b) Cranking inertia as function of $R$. (c)
The estimated $A_2$ during the deformation  as function of $R$.
 }
\label{fig2}
\end{center}
\end{figure}

\begin{figure}
\begin{center}
\resizebox{0.50\textwidth}{!}{
  \includegraphics{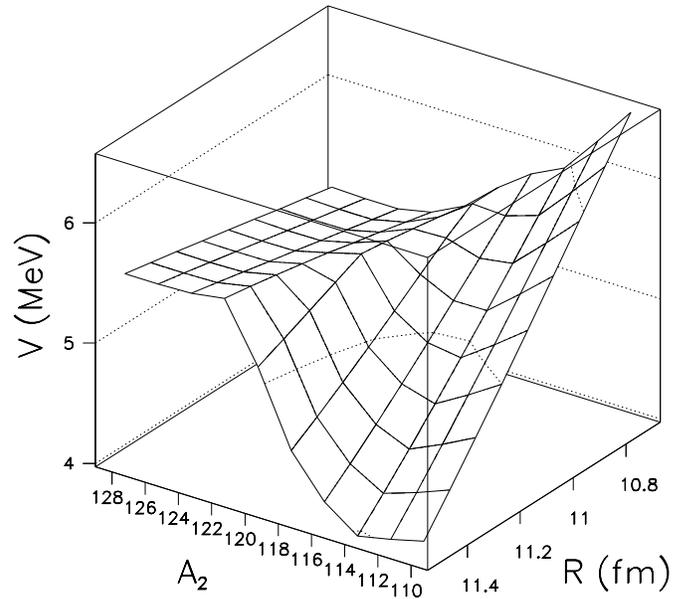}}

\caption{Deformation energy minimized statically with respect the eccentricity
of the second fragment $\varepsilon_2$ and the mass asymmetry parameter $\eta$
as function of $R$ and $A_2$
in the second saddle region. 
 }
\label{fig22}
\end{center}
\end{figure}

\begin{figure}

\begin{center}
\resizebox{0.50\textwidth}{!}{
  \includegraphics{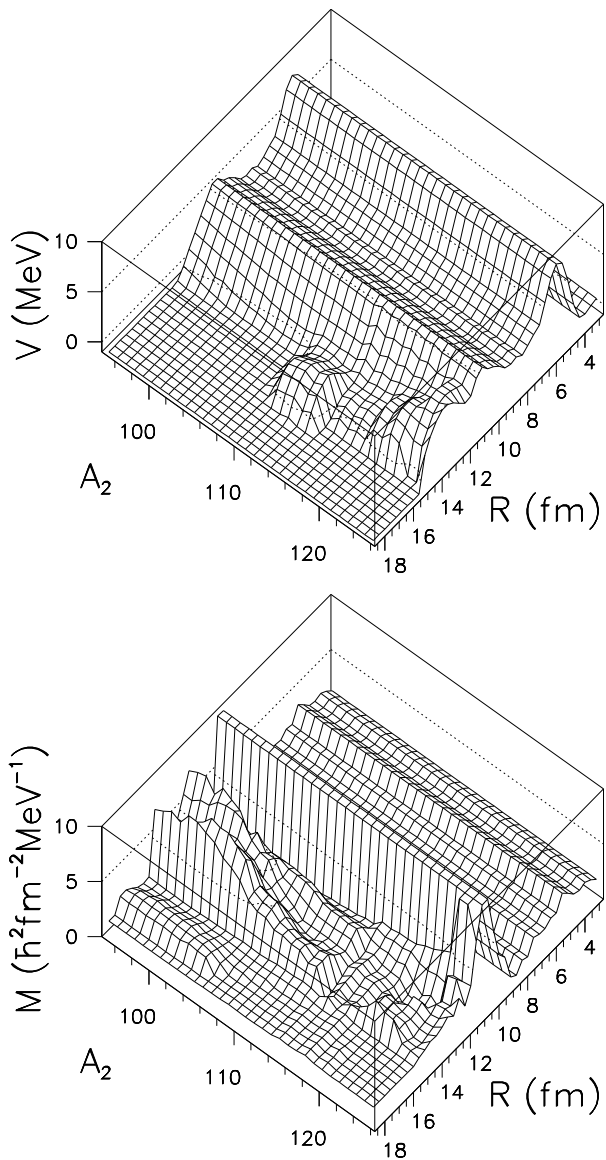}}

\caption{Deformation energy $V$ computed within the microscopic-macroscopic
model for different binary partitions with respect $A_2$ and the elongation $R$.
In the lower plot, the inertia are plotted.
 }
\label{fig3}
\end{center}
\end{figure}

\begin{figure}
\begin{center}
\resizebox{0.50\textwidth}{!}{
  \includegraphics{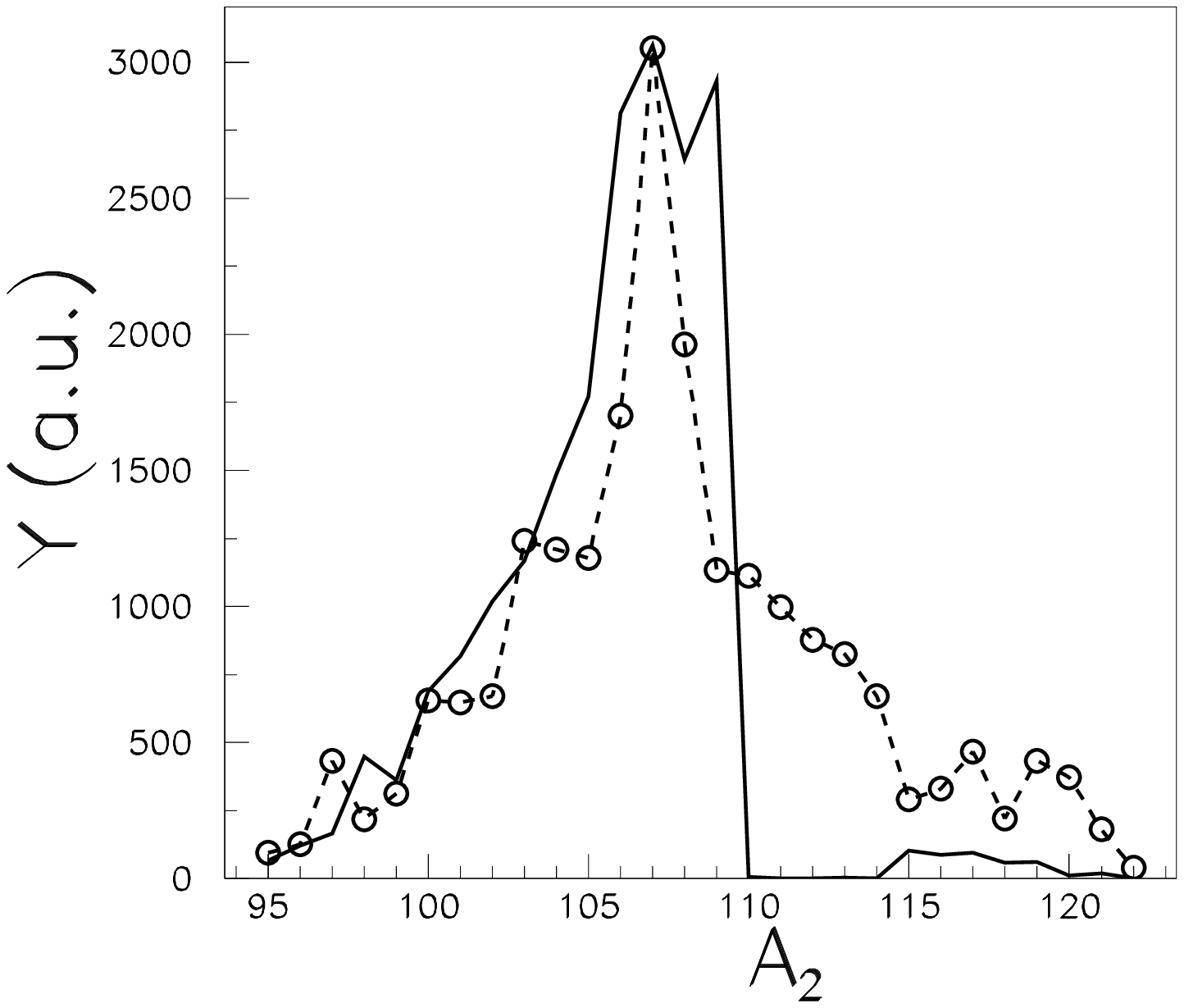}}

\caption{Experimental yields in arbitrary units (dashed line),
compared  with renormalized theoretical penetrabilities calculated 
within the microscopic-macroscopic model (solid line)
with respect to  $A_{2}$.
 }
\label{fig4}
\end{center}
\end{figure}

The penetrability, corresponding to some binary partition, defines the
isotopic yield. This quantity can be 
estimated by using the semiclassical integral \cite{Gam28}
\begin{equation}
\label{penetr}
P_{A_2}=\exp\left\lbrace-2\int_{R_1}^{R_2}
\sqrt{\frac{2M(A_2,R)}{\hbar^2}V(A_2,R)} dR\right\rbrace~,
\label{wkb}
\end{equation}
between internal and external turning points along a fission
path. Two ingredients are
mandatory in order to evaluate the action integral: 
the inertial parameter $M$ and the fragmentation deformation energy 
$V$ (we will call it simply deformation energy). In Eq. (\ref{wkb}),
the inertia $M$ is considered along a given fission path 
 where the main coordinate is the elongation $R$. Therefore,
for this trajectory in the configuration space, the dependencies
of all generalized coordinates versus the main coordinate $R$ are known
and the quantities $V$ and $M$ depend only on the variables $A_2$ and $R$.
The problem of finding the fission trajectory will be treated below.
For a fixed combination $A=A_1+A_2$ the deformation energy has a 
minimum at the charge equilibration point $Z_2$, which we will not 
mention in the following. By definition it is defined as follows
\begin{equation}
\label{poten}
V(A_2,R)=V_N(A_2,R)+V_C(A_2,R)-Q~,
\end{equation}
where $V_N(A_2,R)$ is the nuclear and $V_C(A_2,R)$ Coulomb 
inter-fragment potential. We also introduced the Q-value in terms
of the difference between binding energies of the parent and
the sum of emitted fragments, i.e.
\begin{equation}
\label{Qvalue}
Q=B(Z_1,A_1)+B(Z_2,A_2)-B(Z,A)~.
\end{equation}
For deformed nuclei, due to the fact that the largest emission probability
corresponds to the lowest barrier, the deformation potential decreases 
in the direction of the largest fragment radius.

The deformation energy of the nuclear system 
is the sum between the liquid drop energy $V_{LD}$ and the shell effects 
$\delta E$, including pairing corrections \cite{nix00}, i.e.
\begin{equation}
\label{deform}
V=V_{LD}+\delta E-V_{0}~.
\end{equation}
The energy of the parent nucleus $V_{0}$ is used as a reference
value, so that in the ground state
configuration the deformation energy is zero and asymptotically,
for two separated fragments at infinity, the deformation energy reaches
the minus sum of energies of emitted fragments. Thus,
Eq. (\ref{deform}) coincides with the definition Eq. (\ref{poten}).

The macroscopic energy is obtained within the framework of the
Yukawa-plus-exponential model \cite{davies00}, extended for binary 
systems with different charge densities \cite{poen1,mm01}. 
The Strutinsky prescriptions \cite{brac1} were computed on the basis of 
a new version of the superasymmetric two-center shell model. 
This version solves a Woods-Saxon potential in terms of the two-center 
prescriptions as detailed in Ref. \cite{mirws2}. 
The inertial parameter $M$ is computed  in the framework of the 
cranking model \cite{cranking,crk}. 
We considered only cold fission process. 
Consequently the deformations of the final nuclei are given 
by their ground state values of Ref. \cite{Mol95}.

For comparison with experimental data, the maximal values of the
independent yields for a maximum excitation energy of 7 MeV were
selected from Ref. \cite{ham1}. 
The selected channels address
binary partitions characterized by the following light fragments:
$^{95}$Rb, $^{96}$Rb, $^{97}$Sr, $^{98}$Sr, $^{99}$Y,
$^{100}$Y, $^{101}$Zr, $^{102}$Nb, $^{103}$Zr,
$^{104}$Nb,  $^{105}$Mo, $^{106}$Nb, $^{107}$Mo,
$^{108}$Tc, $^{109}$Mo, $^{110}$Tc, $^{111}$Ru, $^{112}$Rh,
$^{113}$Ru, $^{114}$Rh, $^{115}$Rh, $^{116}$Rh, $^{117}$Pd,
$^{118}$Rh, $^{119}$Pd, $^{120}$Ag, $^{121}$Cd  and $^{122}$Ag.

First of all, the fission path in our five-dimensional 
configuration space must be supplied, that is a dependence between all 
generalized coordinates. This trajectory starts from the ground-state
of the system and reaches the exit point of the barrier.
The ground-state corresponds to the minimal deformation
energy in the first well.
The adiabatic barrier in the multidimensional configuration space
is determined by using the least action principle \cite{brac1}. 
Therefore, a trajectory in the multidimensional space is obtained by 
minimizing the functional (\ref{penetr}) by using a numerical procedure 
as in Ref. \cite{mirws2}. The deformation
energy landscape, minimized versus the parameters $\varepsilon_i$ 
($i$=1,2) and $a_1/a_2$, is plotted in Fig. \ref{fig1} as a
function of the necking coordinate $C$ and the elongation parameter $R$.
In the lower panel of Fig. \ref{fig1} the least action trajectory
is also plotted. In Fig. \ref{fig2}, the adiabatic fission barrier $V$, the
effective mass $M$ and the estimated mass number $A_2$ are plotted
along the fission path. 
According to Eq. (\ref{penetr}) the inertia is computed within the cranking
approximation \cite{cranking} using the expression 
\begin{eqnarray}
M=\sum_{j=1}^{5}\sum_{i=1}^{5}M_{q_{i}q_{j}}{\partial q_{i}\over\partial R}
{\partial q_{j}\over\partial R}
\end{eqnarray}
where $M_{q_{i}q_{j}}$ are the elements of the inertia tensor computed
for the generalized coordinates $q_{i}$ and $q_{j}$.
The second barrier top 
is located at $R$=11 fm and corresponds to  a mass 
$A_2\approx$ 120. 
For a constant charge density, this ratio of the mass asymmetry
addresses a heavy fragment with $A_1-Z_1$=81 and $Z_1$=51, 
these values being close to magic numbers. 
Thus, the second saddle point corresponds to a partition that includes
a double magic fragment.
However, as mentioned in Refs. \cite{boc}, microscopic
approaches to fission \cite{mir,ber,mol} established that the
second saddle point is asymmetrical with a value compatible
with the observed mass ratio of the fragment distribution.
Therefore, the dynamical saddle configuration obtained
within our model is checked by minimizing statically the
deformation energy around $R$=11 fm. The eccentricity $\varepsilon_1$
and the neck parameter $C$ are kept constant. The detailed region
is displayed in Fig. \ref{fig22} confirming that the saddle point is
located at $A_2\approx$120 and $R\approx$11 fm.

Now we are interested in determining the fission barriers that address
all the analyzed partitions. The asymptotic deformations of the
two fragments are taken from the literature \cite{Mol95}.
Thus, the shapes of the initial nucleus up to the second barrier
and those of the final fragments are known.
In order to avoid a complicated determination of the minimal action
path for each partition, a linear variation from initial values of the 
generalized coordinates $\varepsilon_i$ ($i$=1,2) and $a_1/a_2$ 
is postulated starting from the saddle of the
second barrier configuration  up to the 
final ones, characterizing the fragments at the end of the fission process. 

By assuming a final mass asymmetry the values of the deformation energy 
$V$ and the inertia $M$ are plotted in Fig. \ref{fig3} 
as a function of the  elongation $R$ and the light fragment mass $A_2$. 
Some general features of the fission barrier can be extracted.
The shapes of the external barriers change dramatically upon
the mass-asymmetry.
For $A_2 <$110 partitions, a double humped barrier ocurs
while for more symmetric channels ($A_2>110$) a triple humped barrier 
is obtained. Moreover, the exit point from the fission barrier 
is located at lower values of the elongation  $R$ for $A_{2}<110$ than 
for more symmetric channels. The inertia tends to increase when
the mass asymmetry increases. It is clear that the inertia favors the
production of a symmetric partition, that is $A_2>$ 110.

The spontaneous fission half-life is inversely proportional
to the zero point vibration energy and to the probability
of penetration through the fission barrier. If the zero
point vibration energy is considered the same for all partitions,
then the yields are proportional with the penetrability in
each channel.

With these ingredients, the calculated penetrabilities at zero
excitation energies through the  barrier are compared to
experimental yields in Fig. \ref{fig4}. 
We implicitly assume that the penetration of the inner barrier 
is the same for all partitions, so that differences in the 
barrier transmission between channels are induced only by the external 
barrier.

We obtain a very good agreement between the theoretical penetrability 
distribution and the experimental yields for $A_2<110$. 
The maximum theoretical value is at $A_2$=107, while the 
maximum experimental yield is at the same value. A sudden drop of 
theoretical penetrabilities is theoretically obtained for channels
with $A_2>110$. 
As previously noticed, this behavior is connected with the ground state 
shapes of fragments that become oblate for these channels.
Perhaps these oblate shapes are not the best configurations during 
the penetration of the barrier, and the final ground state oblate
configurations are obtained only after the exit from the fission barrier.

Concluding,  we computed the cold fission path in the
potential energy surface of $^{252}$Cf by using the two center
shell model, based on the idea of the cold rearrangements of nucleons 
during the cold fission process.
We obtained a satisfactory agreement with experimental yields,
by considering variable mass and charge asymmetry beyond the first
barrier of the potential surface.
We can see that the mass asymmetry changed from a symmetric to
the asymmetric configuration in the vecinity of the second barrier,
due to the influence of the magic numbers Z=50 and N=82, i.e.
$^{120}_{47}$Ag+$^{132}_{51}$Sb partition.
It was shown that a good agreement with experimental data can be 
obtained only if the fission path proceeds through this saddle
configuration.
The final transmissions depend only on the external barrier
and strongly depend upon the final fragment shapes. 
Due to the fact that prolate shapes are more favorable for the fission 
process, the maximal values of the yields are
shifted towards channels characterized by lower values of $A_2$. 

Thus, the cold fission process of $^{252}$Cf can be called Sn-like 
radioactivity, similar with the Pb-like radioactivity, 
corresponding to various heavy cluster emission processes.
We call all these processes shortly $\kappa$ (cluster) decays.
The peak in the final distribution corresponds to $^{107}$Mo,
due to the mass asymmetry degree of freedom, allowing a lower
barrier from Sn to this nucleus.

\centerline{\bf Acknowledgments}

This work was supported by the contracts IDEI-119 and PN 09 37 01 02 of the
Romanian Ministry of Education and Research.

\end{document}